# Intensity correlation speckles as a technique for removing Doppler broadening

R. Merlin,[1,*] N. Green,[1] I. Szapudi,[2] and G. Tarlé[1]

[1]*The Harrison M. Randall Laboratory of Physics, University of Michigan, Ann Arbor, Michigan 48109-1040, USA*

[2]*Institute for Astronomy, University of Hawaii, 2680 Woodlawn Drive, Honolulu, HI 96822, USA*

A method involving intensity correlation measurements is described, which allows for the complete removal of Doppler broadening in the emission of electromagnetic radiation from far-away sources that are inaccessible to conventional Doppler-free measurements. The technique, relying on a correction to $g^{(2)}$ of order $N^{-1}$, probes the separation between neighboring spectral lines and is also applicable to the elimination of broadening due to collisions ($N$ is the number of emitting particles and $g^{(2)}$ is the second-order field correlation function). Possible applications include a determination of cosmological parameters from red shifts of gravitationally-lensed quasars.

The dependence of the frequency of waves and, in particular, electromagnetic radiation, on the velocity of the source with respect to a detector is known as the Doppler effect. For an ensemble of atoms or molecules in thermal equilibrium, the random motion adds an inhomogeneous term to the natural width of spectral lines, referred to as Doppler broadening, which reflects the spread in velocities. Proven techniques to remove or circumvent Doppler broadening are the early methods of Doppler-free saturated absorption [1] and two-photon spectroscopy [2], and the more recently developed techniques of laser cooling and trapping [3,4]. All these approaches rely on the resonant interaction between a set of counter-propagating laser beams and the radiating particles and, as such, they are ineffective for studying far-away sources, especially astrophysical objects. In this letter, we describe an approach to fully eliminate Doppler broadening, which does not require the manipulation of the sources for it involves solely the detection and processing of spontaneously emitted radiation. The same procedure also serves to remove broadening due to collisions.

The method we propose relates to various intensity- and noise-correlation techniques, the list of which includes one- and two-photon [5] speckle spectroscopy as well as time-domain applications such as fluorescence [6] and photon correlation spectroscopy [7], also known as dynamical light scattering [8]. Speckles contain information about spatial correlations and have thus been used in a variety of applications in microscopy, imaging and studies of surface roughness [9] while time correlations give information on, e.g., the diffusion properties of liquids and small particles in suspension. Closely related to our proposal is the technique of CARS (coherent anti-Stokes Raman scattering) noise-correlation spectroscopy, which purposely uses incoherent light to determine vibrational-resonance differences [10]. In some way, all these methods trace back to the pioneering work of Hanbury-Brown and Twiss who showed that intensity-correlation interferometry allows one to measure the angular sizes of astronomical sources [11]. This and their ensuing work on photon bunching [11] were crucial for understanding the boundary between classical and quantum optics and paved the way for the development of closely related techniques to study collisions in nuclear physics [12].

Consider a conventional chaotic source, such as a gas discharge lamp, involving $M$ spectral lines emitted by $N \gg 1$ identical atoms (or molecules), which radiate incoherently and independently of one another. In their respective center-of-mass rest frame, these lines have frequencies $\{\Omega+\Delta_1,..,\Omega+\Delta_M\}$ where $|\Delta_i| \ll \Omega$ for all $M$ lines. We are interested in the intensity fluctuations of a parallel light beam, which propagates in free space with velocity $c$. We assume that $N \gg 1$ and that the differences $|\Delta_i - \Delta_j|$ ($i \neq j$) are large enough so that collisional and spontaneous-emission broadening can be ignored (collisions are considered later). Chaotic sources are well described by classical theory wherein the total Doppler-broadened complex electric field of a given polarization is given by [13]

$$E = \sum_n E_n(t,z) = \sum_{n=1,N} \sum_{m=1,M} \mathcal{E}_m e^{-i[(\omega_n+\Delta_m)(t-z/c)+\varphi_m^{(n)}]} \quad . \tag{1}$$

Here, $t$ is the time and $z$ is the distance from the atomic cloud's center to the observer; $\omega_n = \Omega(1+v_n/c)$, where $v_n$ is the line-of-sight component of the velocity of the $n$th atom, and $\varphi_m^{(n)}$ is the phase shift of the $n$th atom's $m$th spectral line. Following the classical model, the amplitudes $\{\mathcal{E}_m\}$ are assumed to be deterministic and the same for all the atoms [13]. The $\{\omega_n\}$ and $\{\varphi_m^{(n)}\}$ are taken to be independent and identically distributed random variables that are statistically independent from each other and, respectively, Gaussian (Maxwellian) distributed with mean $\Omega$ and variance $\sigma^2$, and uniformly distributed in $[0, 2\pi]$. Note the fact that, because $|\Delta_i| \ll \Omega$, the Doppler shift $(\Omega+\Delta_m)v_n/c$ is $\approx \Omega v_n/c$ so that Doppler broadening does not significantly affect the separation between lines (this is a key point that made possible the first experimental determination of the Lamb shift [14]). We observe that a full quantum treatment of our problem should yield the same results as those of the classical theory since quantum and classical models give identical predictions for chaotic sources [13].

The intensity-correlation function stems from the product

$$I(t)I(t+\tau) = (c/8\pi)^2 E^*(t)E(t)E^*(t+\tau)E(t+\tau) \qquad (2)$$

where $I$ is the cycle-average intensity at a particular point of detection. In the limit $N \to \infty$, the only terms that survive involve contributions from individual atoms as all other terms vanish because of the random relative phases of different atoms [13]. Let $\langle \ \rangle_\varphi$ denote the statistical ensemble average. Then,

$$\langle I(t)I(t+\tau)\rangle_\varphi = (c/8\pi)^2 \left[ \begin{array}{l} \sum_{n=1,N} \langle E_n^*(t)E_n(t)E_n^*(t+\tau)E_n(t+\tau)\rangle_\varphi \\ + \sum_{n \neq n'} \langle E_n^*(t)E_n(t)\rangle_\varphi \langle E_{n'}^*(t+\tau)E_{n'}(t+\tau)\rangle_\varphi \\ + \sum_{n \neq n'} \langle E_n^*(t)E_n(t+\tau)\rangle_\varphi \langle E_{n'}^*(t+\tau)E_{n'}(t)\rangle_\varphi \end{array} \right]. \qquad (3)$$

Simple calculations give $\langle E_n^*(t)E_n(t)E_n^*(t+\tau)E_n(t+\tau)\rangle_\varphi = \left(\sum_m \mathcal{E}_m^2\right)^2 + \left|\sum_m \mathcal{E}_m^2 e^{-i\Delta_m \tau}\right|^2 - \sum_m \mathcal{E}_m^4$

and $\langle E_n^*(t)E_n(t+\tau)\rangle_\varphi = e^{-i\omega_n \tau} \sum_m \mathcal{E}_m^2 e^{-i\Delta_m \tau}$ (note that the average over the random phases removes the $t$-dependence). Thus, we have exactly

$$\langle I(t)I(t+\tau)\rangle_\varphi = (c/8\pi)^2 \left[ N^2 \left( \sum_{m=1,M} \mathcal{E}_m^2 \right)^2 - N\sum_m \mathcal{E}_m^4 + S(\tau)\left|\sum_{m=1,M} \mathcal{E}_m^2 e^{-i\Delta_m \tau}\right|^2 \right] \qquad (4)$$

with

$$S(\tau) = \sum_{n,n'} e^{-i(\omega_n - \omega_{n'})\tau} = \left|\sum_{n=1,N} e^{-i\omega_n \tau}\right|^2. \qquad (5)$$

$S(\tau)$ can be viewed in the complex plane as the result of a random walk where the step length is unitary and the random directions are Gaussian distributed. The statistical analysis gives

$$\langle S(\tau)\rangle_\omega = N + N(N-1)e^{-\sigma^2\tau^2} \tag{6}$$

where $\langle \ \rangle_\omega$ is the average over the frequencies whereas the variance is

$$\text{Var}_\omega S = \langle S^2(\tau)\rangle_\omega - \langle S(\tau)\rangle_\omega^2 = \\ 8N(N-1)e^{-2\sigma^2\tau^2}[N-1+\cosh\sigma^2\tau^2]\sinh^2(\sigma^2\tau^2/2) \tag{7}$$

Results for a single trial involving $10^4$ atoms are shown in Fig. 1. Note the large, random variations of order $N$ for $\sigma\tau \gtrsim 3$, which reflect the random-walk nature of the sum [15]. For $\sigma\tau \lesssim 3$, fluctuations are negligible and $S(\tau)$ is well represented by its large-$N$ average $N^2 e^{-\sigma^2\tau^2}$. In this range, Eq. (7) gives a standard deviation of order $N^{1.5}$, which explains the fact that, up to $\sigma\tau \approx 3$, departures from the Gaussian behavior are not apparent in the numerical data. The large-$\tau$ fluctuations, with a time scale given by $\sigma^{-1}$, mimic those that occur in the intensity [16] as well as the spatial fluctuations observed in speckle patterns [9]. The single-trial results for $\sigma\tau \gtrsim 3$ are consistent with the approximate expressions $\langle S\rangle_\omega \approx N$ and $\text{Var}_\omega S \approx N^2$, which are valid in the limit $Ne^{-\sigma^2\tau^2} \ll 1$.

The normalized intensity correlation function, also known as the degree of second-order temporal coherence is defined as

$$g^{(2)}(\tau) = \langle I(t)I(t+\tau)\rangle_t / I_0^2 \tag{8}$$

where $\langle f(t)\rangle_t = T^{-1}\int_T f(t)dt$ and $I_0 = \langle I(t)\rangle_t$ is the average intensity. The integration interval $T$ is taken to be much longer than the characteristic time of the intensity fluctuations, $\sigma^{-1}$, so that the time average samples all of the values consistent with the ergodic properties of the source. With this choice, the average over time, relevant to experiments, is equivalent to the statistical average over the phases and frequencies. Thus, we get from Eqs. (4) and (6)

$$g^{(2)}(\tau) = 1 - \frac{1}{N} \frac{\sum_m \mathcal{E}_m^4}{\left(\sum_{m=1,M} \mathcal{E}_m^2\right)^2} + \left|\frac{\sum_{m=1,M} \mathcal{E}_m^2 e^{-i\Delta_m \tau}}{\sum_{m=1,M} \mathcal{E}_m^2}\right|^2 \frac{\langle S(\tau) \rangle_\omega}{N^2} \quad . \tag{9}$$

Since $\langle E^*(t)E(t+\tau)\rangle_t = \langle \sum_n e^{-i\omega_n \tau}\rangle_\omega \sum_m \mathcal{E}_m^2 e^{-i\Delta_m \tau}$, this expression becomes identical to that from the Siegert relation $g^{(2)} = 1 + |g^{(1)}|^2$ [17] for $N \gg 1$ ($g^{(1)}(\tau) = \langle E_n^*(t)E_n(t+\tau)\rangle_t / I_0$ is the first-order temporal coherence).

In the following, we address the question as to how to extract information on the separation between lines from measurements of $g^{(2)}$. In situations where the Doppler broadening is not significant, that is, for $\sigma \lesssim \min|\Delta_i - \Delta_j|$ ($i \neq j$), the line separations can be obtained from the behavior of $g^{(2)}$ for $\tau \lesssim 2\pi / \min|\Delta_i - \Delta_j|$. In such cases or, more generally, for $Ne^{-\sigma^2 \tau^2} \gg 1$

$$g^{(2)}(\tau) \approx 1 + \left|\frac{\sum_{m=1,M} \mathcal{E}_m^2 e^{-i\Delta_m \tau}}{\sum_{m=1,M} \mathcal{E}_m^2}\right|^2 e^{-\sigma^2 \tau^2} \quad . \tag{10}$$

Thus, the frequency differences can be gained by Fourier transforming $g^{(2)}(\tau)$. Since the separations can more easily be obtained using a conventional spectrometer, however, these cases are not very interesting experimentally. Further, it is also clear that the Gaussian factor makes it essentially impossible to probe the range $\tau \lesssim 2\pi / \min|\Delta_i - \Delta_j|$ for $\sigma \gtrsim \max|\Delta_i - \Delta_j|$.

To resolve neighboring lines in systems with large broadening, we consider instead the limit $Ne^{-\sigma^2 \tau^2} \ll 1$ (which necessarily implies $\sigma\tau \gg 1$ and $\langle S(\tau)\rangle_\omega \approx N$) and get

$$g^{(2)}(\tau) \approx 1 - \frac{1}{N} \frac{\sum_m \mathcal{E}_m^4}{\left(\sum_{m=1,M} \mathcal{E}_m^2\right)^2} + \frac{1}{N}\left|\frac{\sum_{m=1,M} \mathcal{E}_m^2 e^{-i\Delta_m \tau}}{\sum_{m=1,M} \mathcal{E}_m^2}\right|^2 \quad . \tag{11}$$

Note that $g^{(2)}(\tau) \approx 1 - N^{-1}/2 + N^{-1}\cos^2(\delta\tau/2)$ for a doublet with $\mathcal{E}_1 = \mathcal{E}_2$; $\delta = \Delta_1 - \Delta_2$ is the spacing between the two lines. Once again, a simple Fourier transform can be used to obtain the line separations. Since the average and the standard deviation of $S(\tau)$ are of the same order, the signal-to-noise ratio can be enhanced by first filtering the data through a multiplicative-noise removal algorithm [18] of the sort commonly used in image processing [19].

Equation (11) is the main result of our work. It shows that intensity correlation measurements can be used to resolve arbitrarily close neighboring lines in the presence of arbitrarily large Doppler broadening. Concerning practical implementations, two comments are in order. First, we note that all the $\tau$-independent contributions to $g^{(2)}$ and, in particular, the one derived from the dominant background term of order $N^2$ in Eq. (4) can be experimentally eliminated if one uses a shaker controlled by a lock-in amplifier [20]. Second, the condition $Ne^{-\sigma^2\tau^2} \ll 1$ or, alternatively, $\ln N \lesssim (2\pi\sigma/\max|\Delta_i - \Delta_j|)^2$, is easily satisfied in gas discharge tubes since typical densities are in the range $10^{16}$-$10^{17}$ cm$^{-3}$. As an example, consider measuring the Lamb shift in hydrogen (~ 1 GHz) using the Balmer $\alpha$ spectrum. For a Doppler width of 2 GHz, we find that this condition is met for $N \ll 10^{68}$.

The method we propose also serves to remove collision broadening, which is accounted for by a classical field of the form

$$E = \sum_n E_n(t) = \sum_{n=1,N} \sum_{m=1,M} \mathcal{E}_m e^{-i[(\omega_0 + \Delta_m)(t-z/c) + \varphi_m^n(t)]} \quad (12)$$

involving time-dependent phases that change randomly and suddenly when a collision occurs [13]. Using a procedure similar to that for Doppler broadening, we find

$$g^{(2)}(\tau) = 1 - \frac{1}{N}\frac{\sum_m \mathcal{E}_m^4}{\left(\sum_{m=1,M}\mathcal{E}_m^2\right)^2} + \frac{1}{N^2}\frac{\left\langle\left|\sum_{n=1,N,m=1,M} e^{i[\varphi_m^n(t+\tau)-\varphi_m^n(t)]}|\mathcal{E}_m|^2 e^{i\Delta_m\tau}\right|^2\right\rangle_t}{\left(\sum_{m=1,M}|\mathcal{E}_m|^2\right)^2} . \qquad (13)$$

It is easy to see that this expression satisfies the Siegert relation for $N \gg 1$. Moreover, since the phases become uncorrelated at times much longer than the collision time $\tau_C$, it can be shown that the above expression becomes identical to Eq. (11) for $Ne^{-\tau^2/\tau_C^2} \ll 1$, provided collisions do not change the relative phase of neighboring lines. Therefore, the discussion of the previous paragraphs applies also to collision broadening.

In conclusion, we have shown that frequency differences that are much smaller than the Doppler or the collision width can be determined using intensity-correlation methods by uncovering a critical correction to $g^{(2)}$ of order $N^{-1}$. Our approach holds promise for resolving fine spectral features in situations where gaining access to or disturbing the source is, respectively, impracticable or undesirable. One such a case bears on the determination of the expansion of the universe from the red shifts of a distant galaxy or quasar, measured at different epochs, as proposed by Sandage and Loeb [21,22,23]. To that end, multiply imaged quasars, a result of gravitational lensing, present an even better opportunity since light from the same object follows paths of different lengths, resulting in effective time delays as large as 100 years [24]. For an acceleration of about 2.5 cm/s-year, the expected frequency shift of the Ly-α line of hydrogen is several orders-of-magnitude smaller than the Doppler broadening and, thus, beyond the reach of a conventional spectrograph. The results presented here open up the possibility that expansion-induced shifts could be determined from measurements of the intensity correlation between multiple images.

The authors are grateful to the anonymous referee who provided very useful and detailed comments on an earlier version of the manuscript.

# REFERENCES

* Corresponding author. merlin@umich.edu

FIGURE CAPTIONS

Figure 1 – Single-trial computer simulation of $S(\tau) = \left|\sum_m e^{-i\omega_m \tau}\right|^2$ (black curve, logarithmic scale). The horizontal orange line is $N = 10^4$ and the dashed gray curve is $N^2 e^{-\sigma^2 \tau^2}$. Inset: $S(\tau)/N$ vs. $\tau$ (linear scale).

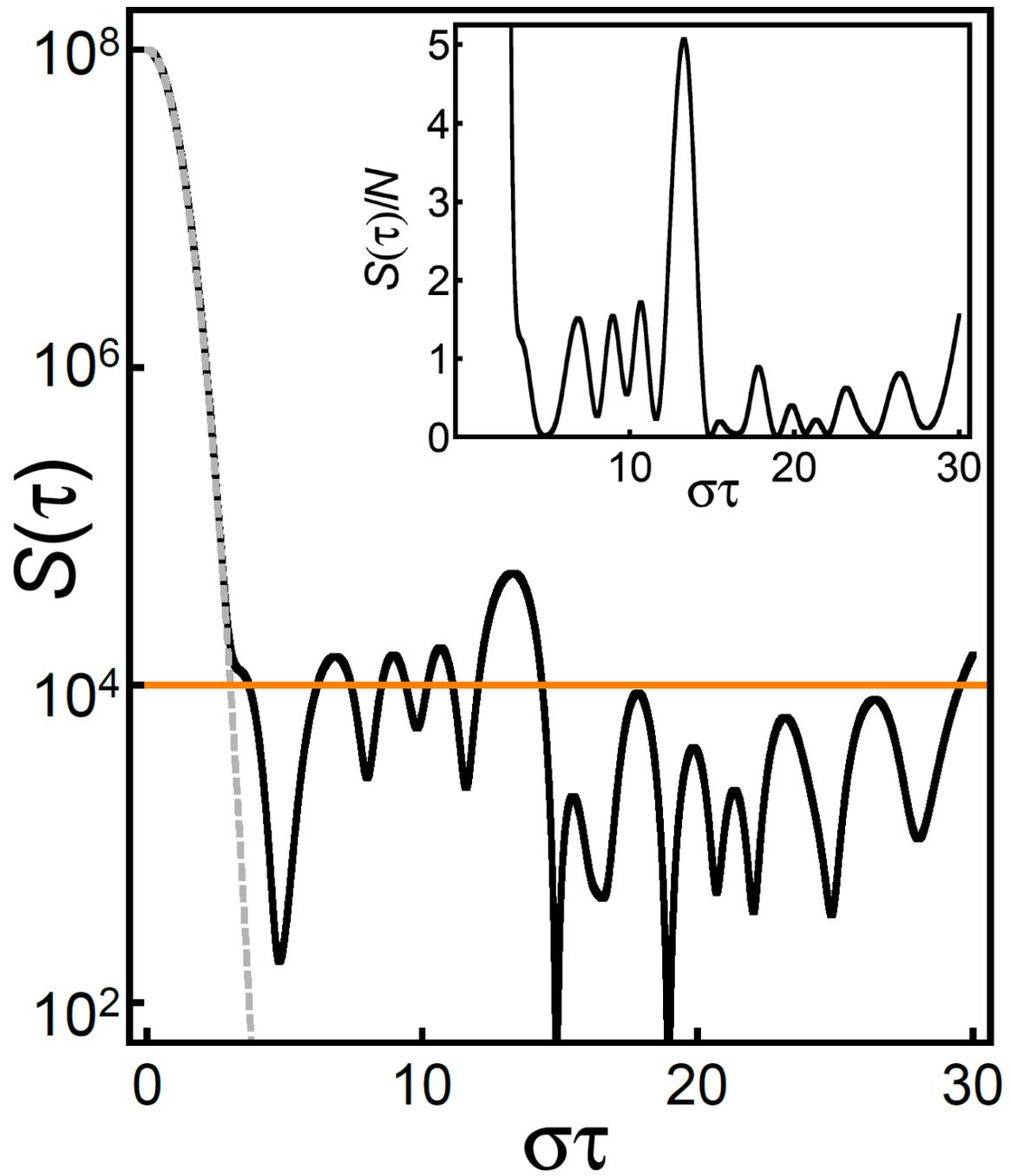

FIGURE 1